\documentclass[prb,showpacs,twocolumn]{revtex4}
\usepackage{psfrag}
\usepackage{graphicx}
\usepackage{amsmath}
\usepackage{amssymb}
\usepackage[caption=false]{subfig}

\newcommand{\qpar}{$q$ }
\newcommand{\mr}{\mathrm}
\newcommand{\mb}{\mathbf}
\newcommand{\VS}{$\mathrm{VS}_2$}

\newcommand{\NbS}{$\mathrm{NbS}_2$}

\newcommand{\TaSe}{$\mathrm{TaSe}_2$}
\begin{document}

\title{Plasmons in metallic mono- and bilayer transition metal dichalcogenides}

\author{Kirsten Andersen}
\email{kiran@fysik.dtu.dk}
\affiliation{Center for Atomic-scale Materials Design (CAMD) and Center for Nanostructured Graphene (CNG), Department of
Physics \\ Technical University of Denmark, DK - 2800 Kgs. Lyngby, Denmark}

\author{Kristian S. Thygesen}
\email{thygesen@fysik.dtu.dk}
\affiliation{Center for Atomic-scale Materials Design (CAMD) and Center for Nanostructured Graphene (CNG), Department of
Physics \\ Technical University of Denmark, DK - 2800 Kgs. Lyngby, Denmark}

\begin{abstract}
  We study the collective electronic excitations in metallic single-
  and bilayer transition metal dichalcogenides (TMDCs) using time
  dependent density functional theory in the random phase
  approximation. For very small momentum transfers (below
  $q\approx0.02$~\AA$^{-1}$) the plasmon dispersion follows the
  $\sqrt{q}$ behavior expected for free electrons in two dimensions.
  For larger momentum transfer the plasmon energy is significantly red
  shifted due to screening by interband transitions. At around
  $q\approx 0.1$ \AA$^{-1}$ the plasmon enters the dissipative
  electron-hole continuum and the plasmon dispersions flatten
  out at an energy around 0.6-1.1 eV, depending on the material.  Using bilayer
  NbSe$_2$ as example, we show that the plasmon modes of a bilayer
  structure take the form of symmetric and anti-symmetric hybrids of
  the single-layer modes. The spatially anti-symmetric mode is rather
  weak with a linear dispersion tending to zero for $q=0$ while the
  energy of the symmetric mode follows the single-layer mode
  dispersion with a slight blue shift.
\end{abstract}

\maketitle
\section{Introduction}
The success of graphene research has created a surge in interest for
other types of atomically thin two-dimensional (2D)
materials.\cite{britnell,strano}. A particularly interesting class of
2D materials are the transition metal dichalcogenides (TMDC) whose
electronic properties range from semiconducting to metallic and even
superconducting.  In combination with graphene and 2D insulators like
hexagonal boron-nitride, the TMDCs could form the basis for
artificially layered van der Waals structures with tailored electronic
properties. While monolayers of semiconducting TMDCs have been
fabricated and studied quite
extensively\cite{radisavljedic,mak,splendiani}, metallic TMDCs have
received less attention, possible due to stability issues\cite{geim}.
However, exfoliation of atomically thin layers of TaSe$_2$, TaS$_2$,
and NbSe$_2$ have been reported\cite{castellanos,ayari,staley}.

The optical properties of a metallic system are to large extent governed
by the collective excitations known as plasmons.  In (doped) graphene,
it was demonstrated that the energy of the so-called metallic plasmon,
which is formed by intraband transitions within the $\pi$ or $\pi^*$
bands, can be varied by electrostatic gating or nanostructuring.
Furthermore, the plasmons facilitates a strong spatial confinement of
light making graphene interesting for applications within
plasmonics\cite{grapheneplas1,grapheneplas2}.  However, the energy of
the metallic plasmon is restricted by the achievable carrier
concentration and this limits applications of graphene plasmonics to
the tetrahertz regime.  Metallic TMDCs have much higher charge carrier
densities leading to plasmon energies of around 1 eV in bulk
TMDCs\cite{liang,konig_plasmon_2012,cudazzo_plasmon_2012,faraggi2012NbSe2}.
Together with the possibility of tuning the plasmon energies and
lifetimes through quantum confinement or plasmon
hybridization\cite{Prodan_2003,Wang_2006}, this makes few-layer TMDCs
interesting candidates for nano-plasmonic applications in the optical
frequency regime. We mention that the efficient coupling of 2D plasmons
and light generally requires some kind of scattering of the
light to overcome the momentum mismatch\cite{mortensen}.

First-principles calculations of the $q=0$ loss spectrum of several
single- and bilayer TMDCs was recently reported in Ref.
\onlinecite{johari}.  While these calculations revealed the energy of
the $\pi$ and $\pi+\sigma$ interband plasmons they did not address the
nature of the intraband plasmons of the metallic systems which are
only seen for finite $q$ due to the $\sqrt{q}$ dispersion in the small
$q$ limit.

In this paper we study the intraband plasmons in monolayers of six
different metallic TMDCs using first-principles calculations. For very
small momentum transfer ($q<0.02$ \AA), the plasmon energy follows the
classical $\sqrt{q}$ behavior of free electrons in 2D. At larger $q$
the plasmons become screened by interband transitions leading to
significant redshift compared to the free electron result, and the
plasmon dispersions flatten out at an energy in the range 0.6-1.1 eV, depending
on the material. At around $q=0.1$ \AA$^{-1}$, the plasmon enters the
dissipative region of interband transitions and the plasmons acquire a
finite lifetime.  Using a recently developed method for visualizing
the induced density associated with a plasmon mode, we study how the
single layer plasmons hybridize to form a symmetric and a (weak)
anti-symmetric plasmon in a bilayer structure. The symmetric mode of
the bilayer is slightly blue shifted relative to the single-layer plasmon. This
pushes it closer towards the dissipative interband continuum and
reduces its strength compared to the single-layer plasmon.

\section{Method}
All density functional theory (DFT) calculations have been performed with the GPAW electronic
structure code\cite{GPAW2}.  We study single layers of TMDCs with the
chemical form $\mr{MX}_2$, where $\mr{M}$ belongs to the group $5$
transition metals $\mr{V}$, $\mr{Nb}$ and $\mr{Ta}$ and the chalcogen
$\mr{X}$ is either $\mr{S}$ or $\mr{Se}$, which results in six
different combinations. The layers were modeled within a periodic
supercells with $20$ \AA\ of vacuum separating the periodic images in
the perpendicular direction. The structures were relaxed with the Perdew-Burke-Ernzerhof (PBE)
exchange-correlation functional\cite{pbe}. The single-particle states used to construct the
response function were calculated using the orbital-dependent GLLBSC (Gritsenko, Leeuwen, Lenthe,
and Baerends potential\cite{gritsenko} with the modifications from Kuisma
\emph{et al.}\cite{GLLBSC}) functional which
has been shown to give accurate band energies of
semiconductors. In the context of plasmons, correct band positions are important to describe the onset of interband transitions correctly. The 2D Brillouin zone (BZ) was sampled using $64\times 64$ k-points for most calculations while a dense grid of
$256\times 256$ was used in order to study very small momentum transfers.

The dielectric matrix for in-plane wave vectors $\mb q$ was calculated in the random phase approximation (RPA)\cite{bohm_pines} following the implementation of Ref. \onlinecite{Yan_2011} 
\begin{equation}\label{eq:eps}
 \epsilon_{\mb{G},\mb{G}'}(\mb{q},\omega) = \delta_{\mb{G},\mb{G}'} - \frac{4\pi}{|\mb{q}+\mb{G}|^2} \chi^0_{\mb{G},\mb{G}'}(\mb{q},\omega).
\end{equation}
For the reciprocal lattice vectors $\mb G$ and $\mb G'$ we used a cutoff of
50 eV to account for local field effects, and empty states up to 30 eV above the Fermi level were
included in the non-interacting response function matrix $\chi^0_{\mb{G},\mb{G}'}(\mb{q},\omega)$. We have checked that the plasmon energies are well converged (to within 0.1 eV) with these parameters. We note that inclusion of the ALDA kernel was found to have negligible effect on the loss spectrum of bulk NbSe$_2$\cite{cudazzo_plasmon_2012}, and based on this we expect the RPA to be valid also for the 2D TMDCs studied here. In order to remove
the interaction between supercells, the Coulomb interaction was
truncated in the direction perpendicular to the
layers\cite{rozzi2006}. If this is not done the plasmons in neighboring layers will interact for in-plane wave vectors $q<2\pi/L$, where $L$ is the distance between the layers. In particular the plasmon energy will incorrectly tend to a finite energy in the $q\to 0$ limit. 

The electron energy loss spectrum (EELS) was calculated from the inverse of the macroscopic dielectric function, $\epsilon_M(\mb{q},\omega)=1/\epsilon_{\mb{G}=\mb{G}=0}^{-1}(\mb{q}, \omega)$: 
 \begin{equation}\label{eq.eels}
\mr{EELS}(\mb{q},\omega) = -\mr{Im}\epsilon_{M}^{-1}(\mb{q}, \omega).
 \end{equation}
The plasmon energies are obtained as the peaks in the loss spectrum which is calculated for different momentum transfers parallel to the layer, $\mb{q}$, and the integrated weight of the loss peaks is used as a measure of the strength of the plasmon mode.

\begin{figure*}
\begin{center}
  \includegraphics[width=1.0\linewidth]{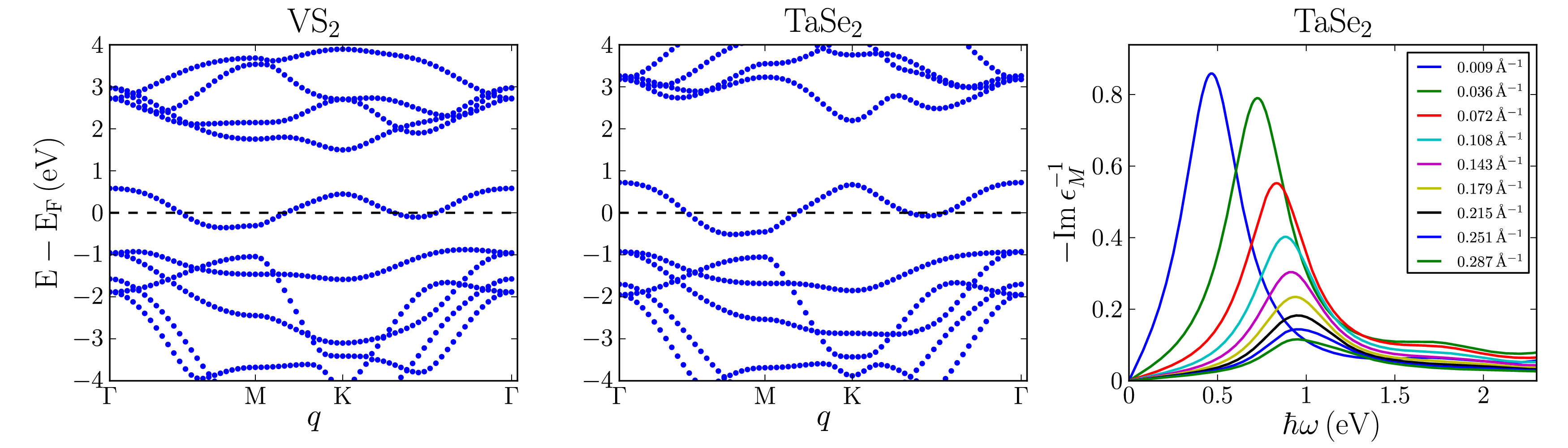}
\caption{Left and middle panel: Band structure of the transition metal dichalcogenides \VS and \TaSe. The metallic band is seen to be separated from all other bands. Right panel: Electron energy loss spectrum (EELS) calculated for \TaSe for increasing momentum transfers, \qpar, along the $\Gamma-M$ direction. }
\label{fig:bands}
\end{center}
\end{figure*}

\subsection{Plasmon eigenmodes}\label{sec.plasmon}
A simple method to identify and compute the plasmon modes of a metallic nanostructure was introduced in Ref. \onlinecite{Andersen_2012}. In the present work, the method has been used to analyze the plasmon excitations of the bilayer structure. An outline of the method is given below.  

A self-sustained charge density oscillation, $\rho(\omega,\mathbf r)$, can exist in a material if the related potential, satisfying Poisson's equation $\nabla^2 \phi(\omega,\mb r)=-4\pi \rho(\omega,\mb r)$, obeys the equation
\begin{equation}
\int \epsilon(\omega,\mathbf r,\mathbf r')\phi(\omega,\mathbf r')d\mathbf r'=0
\end{equation} 
This corresponds to the case of having a finite induced potential
in the absence of an external potential, which is the criterion
for the existence of a plasmon. In general, this
equation cannot be exactly satisfied due to a finite imaginary
part originating from single-particle transitions, which will
lead to damping of the charge oscillation. When the damping
in the material is small, it is sufficient to require only that the
real part of $\epsilon$ vanishes and use the following definition for the
potential associated with a plasmon mode of frequency $\omega_n$:
\begin{equation}
\int \epsilon(\omega_n,\mathbf r,\mathbf r')\phi_n(\omega_n,\mathbf r')d\mathbf r'=i\Gamma_n \phi_n(\omega_n,\mathbf r),
\end{equation} 
where $\Gamma_n$ is a real number. The plasmon modes are thus
eigenfunctions corresponding to purely imaginary eigenvalues of the dielectric function. Physically, they represent the potential
associated with self-sustained charge-density oscillations
damped by electron-hole pair formations at the rate $\Gamma_n$. The
right-hand-side eigenfunctions, $\phi_n$, define the induced
potential of the eigenmodes, and the corresponding induced
density $\rho_n$ can be obtained as the left-hand-side (dual)
eigenfunctions of the dielectric function\cite{Andersen_2012}. In the case of larger and frequency-dependent damping, a more accurate approach
is to use the eigenvalues of $\epsilon(\omega,\mathbf r,\mathbf r')$, denoted $\epsilon_n(\omega)$, and use
the criterion
\begin{equation}\label{eq.eels2}
-\text{Im}\epsilon_n(\omega)^{-1}\text{ is a maximum}
\end{equation}
to define a plasmon mode. This definition takes into
account that a finite imaginary part of $\epsilon$ may shift the plasmon
peaks in the loss spectrum away from the zeros of $\text{Re}\epsilon(\omega)$. 

In contrast to the standard definition of the loss spectrum, Eq. (\ref{eq.eels}), which measures the energy loss of a plane wave perturbation, the definition (\ref{eq.eels2}) provides a resolution of the loss spectrum into the intrinsic plasmon eigenmodes of the system. As a simple example we mention the resolution of the loss spectrum of a thin metal film into bulk and surface modes\cite{Andersen_2012}.
 
In practice, the plasmon eigenmodes are obtained by diagonalizing the dynamical dielectric function (in the $\mb{G}$ and $\mb{G}'$ basis) for each point on a frequency grid. The dielectric eigenvalue spectrum
results in a set of distinct eigenvalue curves that evolves
smoothly with energy, each curve corresponding to a separate
plasmon eigenmode, see Fig. 4 (left panel).

\section{Results}
In the left panels of Fig.~\ref{fig:bands} we show the band structures for VS$_2$ and TaSe$_2$ obtained with the GLLBSC functional. We note that the band structures obtained with PBE are quite similar.
Compared to
semiconducting TMDCs such as $\mathrm{MoS}_2$, the reduced $d$-state
occupation of the transition metal leads to a half-filled band. From the band structure plotted in Fig.~\ref{fig:bands} along the high symmetry $\Gamma-M$ direction it can be seen that the metallic band crosses the Fermi level along two curves in the 2D Brillouin zone. The
band is situated in a gap of approximately 2.5-3 eV and stems from
hybridization of the $d$-state of the transition metal with the vertical
$p_z$ orbitals of the chalcogen. The band structures of the six
materials are very similar, showing only minor variations in the width
of the metallic band, and details in the position of the fully
occupied or unoccupied bands. A comprehensive study of the electronic structure of several 2D TMDCs was reported in \onlinecite{ataca}.

\begin{figure}
\begin{center}
  \includegraphics[width=\linewidth]{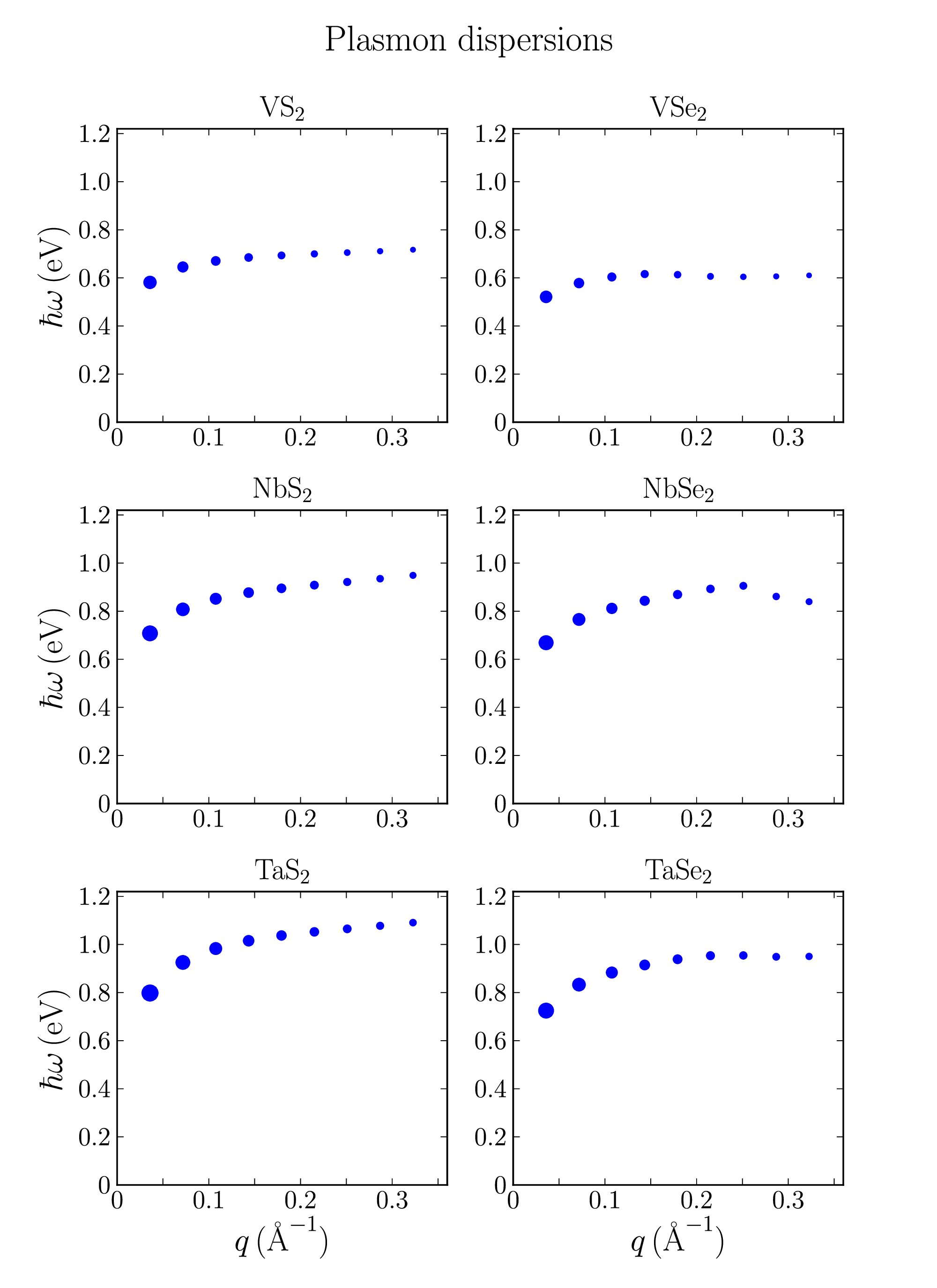}
\caption{Plasmon dispersion of the six TMDCs along the $\Gamma-\mr{M}$
direction, obtained from the peaks in the loss spectra. The size of the markers represents the strength of the plasmon defined as the area under the loss peak. The energy dispersion tends towards zero for low momentum transfers as expected for a metallic 2D plasmon. At larger \qpar however, the dispersion is constant or in some case negative, and the strength of the plasmon is significantly reduced. }
\label{fig:disp}
\end{center}
\end{figure}

\begin{figure}
\begin{center}
  \includegraphics[width=0.98\linewidth]{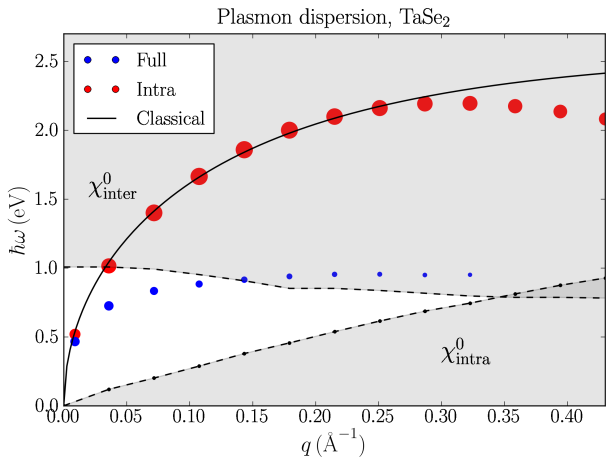}
\caption{Plasmon dispersion along the $\Gamma-\mr{M}$
direction calculated from the full response function (red dots) and from a response function including only intra-band transitions (blue dots). The result of a classical model for the plasmons in a 2D film is shown for comparison. The shaded areas indicate the dissipative regions of Landau damping by inter- and intra-band transitions, respectively.}
\label{fig:tase2}
\end{center}
\end{figure}

An example of an electron energy loss spectrum is shown in
Fig.~\ref{fig:bands}, here calculated for \TaSe\ for momentum
transfers in the $\Gamma-\mr{M}$
direction. Momentum transfers along $\Gamma-\mathrm{K}$ and $\Gamma-\mathrm{K}$ was found to produce very similar results. The pronounced peaks in the loss spectrum are due to plasmon excitations. The plasmon
dispersions of the six TMDCs are shown in Fig.~\ref{fig:disp}, where
the strength of the resonances is indicated by the size of the
markers. In general, higher plasmon energies (for fixed momentum transfer) are found for the TMDCs with heavier transition metal ions. On the other hand, the TMDCs containing heavier
species have larger lattice constants and thus lower density of free
charge carriers. The trend in plasmon energy is therefore opposite to the free electron model
wich predicts the plasmon energy to scale with the charge density as
$\omega_P\propto \sqrt{n}$. As we show below, the TMDC plasmon dispersions deviate from the 2D free-electron model due to (i) the non-parabolic shape of the metallic band, and (ii) screening by interband transitions. We note that the $\sqrt{q}$-behavior of 2D free-electron plasmons in the small $q$ limit is not directly seen from the figure. This is simply because the applied $q$-point grid is not fine enough to reveal this behavior around $q=0$ (this is evident from Fig. \ref{fig:tase2} where a finer $q$-point mesh was used).
We mention that crystal local field effects within the plane may be a third reason for deviations with the 2D free-electron model. However, the homogeneous nature of the 2D TMDC suggests that this will be a rather weak effect.

To investigate the origin of the deviation from free-electron behavior we have studied the separate effects of intra- and interband transitions on the plasmon resonances of mono-layer \TaSe. The non-interacting density response function can be divided into two contributions:
\begin{equation}
\chi^0_{\mb{G},\mb{G}'}(\mb{q},\omega) = \chi^{0, \mr{intra}}_{\mb{G},\mb{G}'}(\mb{q},\omega) + \chi^{0, \mr{inter}}_{\mb{G},\mb{G}'}(\mb{q},\omega),
\end{equation}
where the first term is obtained by only including transitions within the metallic band in the response calculation, and the second term by including all transitions between separate bands. When the dielectric matrix is calculated from $\chi^{0, \mr{intra}}_{\mb{G},\mb{G}'}(\mb{q},\omega)$ using Eq.~(\ref{eq:eps}), 
the pure intraband plasmon, i.e. neglecting screening by the inter-band transitions, is obtained. The result is shown in Fig.~\ref{fig:tase2}, where the dispersion of the pure intraband plasmon is plotted together with the dispersion obtained from the full calculation. The dispersion of the intraband plasmon is furthermore compared to a classical free-electron model for the 2D plasmon of a thin film\cite{ferrell}: 
\begin{equation}
\omega = \frac{\omega_\mr{P}}{\sqrt{2}} \sqrt{1- e^{-q d}},
\end{equation}
where $d$ is the thickness of the film. The bulk plasmon frequency, $\omega_P$, can be obtained from the free electron density, $n$, originating from the half-filled metallic band 
and the effective mass of the electron, $m^*$, through the relation: $\omega_\mr{P}^2= \frac{ne^2}{m^* \epsilon_0}$. For the film thickness, $d$, which is also used to define the free-electron density, we use $d=\frac{3}{2} d_{\mr{Se}-\mr{Se}}$, where $d_{\mr{Se}-\mr{Se}}$ is the distance between the two Se layers. This number agrees well with the \emph{ab-initio} result for the width of the ground state electron density of the layer, however, the result of the classical model is relatively insensitive to variations in this parameter: if $d$ is changed by 10\% the plasmon energy $\omega_P$ changes by around 5\%. A good fit to the intraband \emph{ab-initio} results in the small $q$ region is obtained when the effective mass is set to $m^*= 2 m_e$ (see Fig. \ref{fig:tase2}). We note that this value for the effective mass agrees reasonable well with the effective mass derived directly from the \emph{ab-initio} band structure which lie in the range $1m_e-3m_e$ depending on where along the Fermi curve it is evaluated. At higher $q$, the finite width of the metallic band (see Fig. 1) leads to deviations between the \emph{ab-initio} intraband plasmon and the free-electron result with the former showing a negative dispersion. 

\begin{figure*}
\begin{center}
  \includegraphics[height=0.28\linewidth]{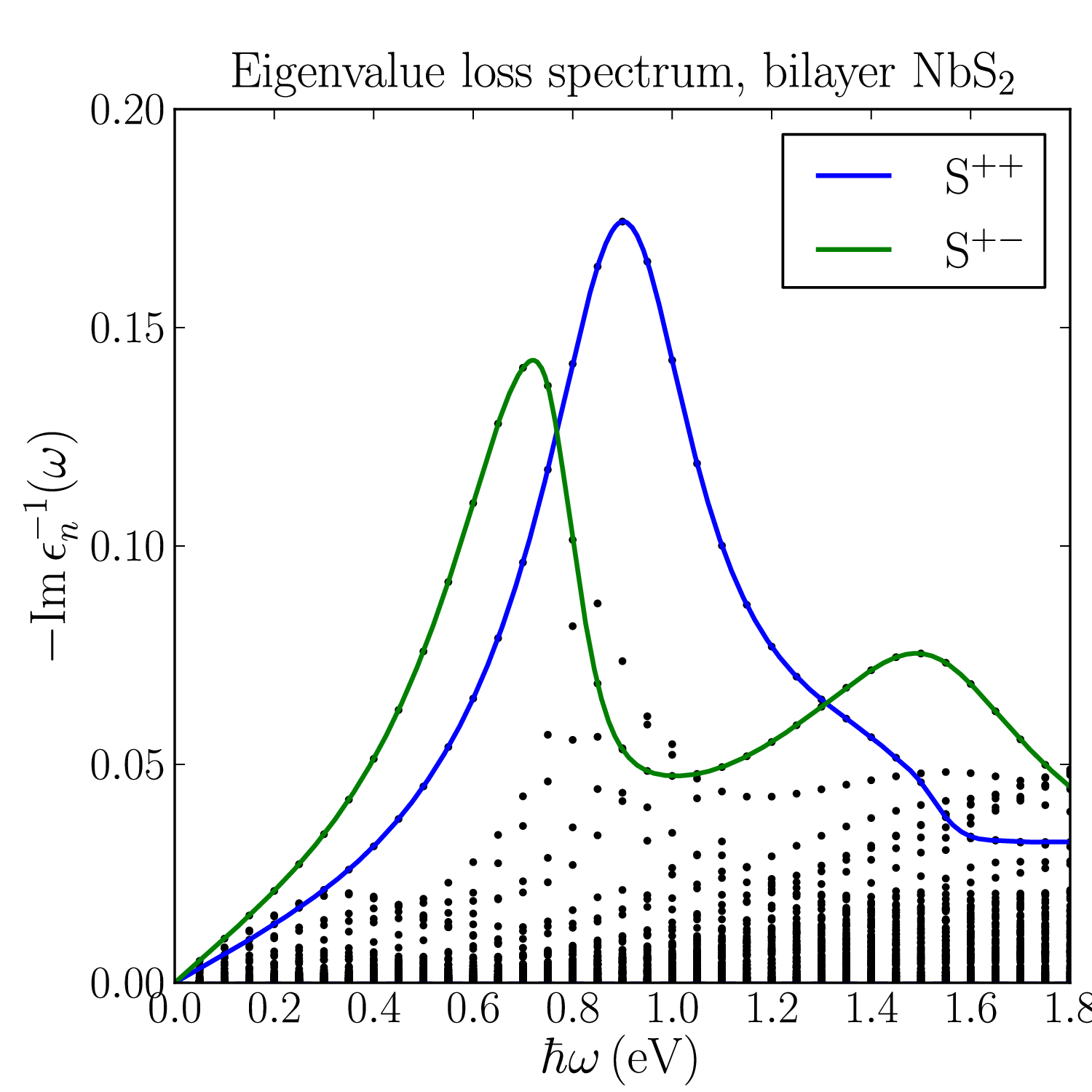}
   \includegraphics[height=0.28\linewidth]{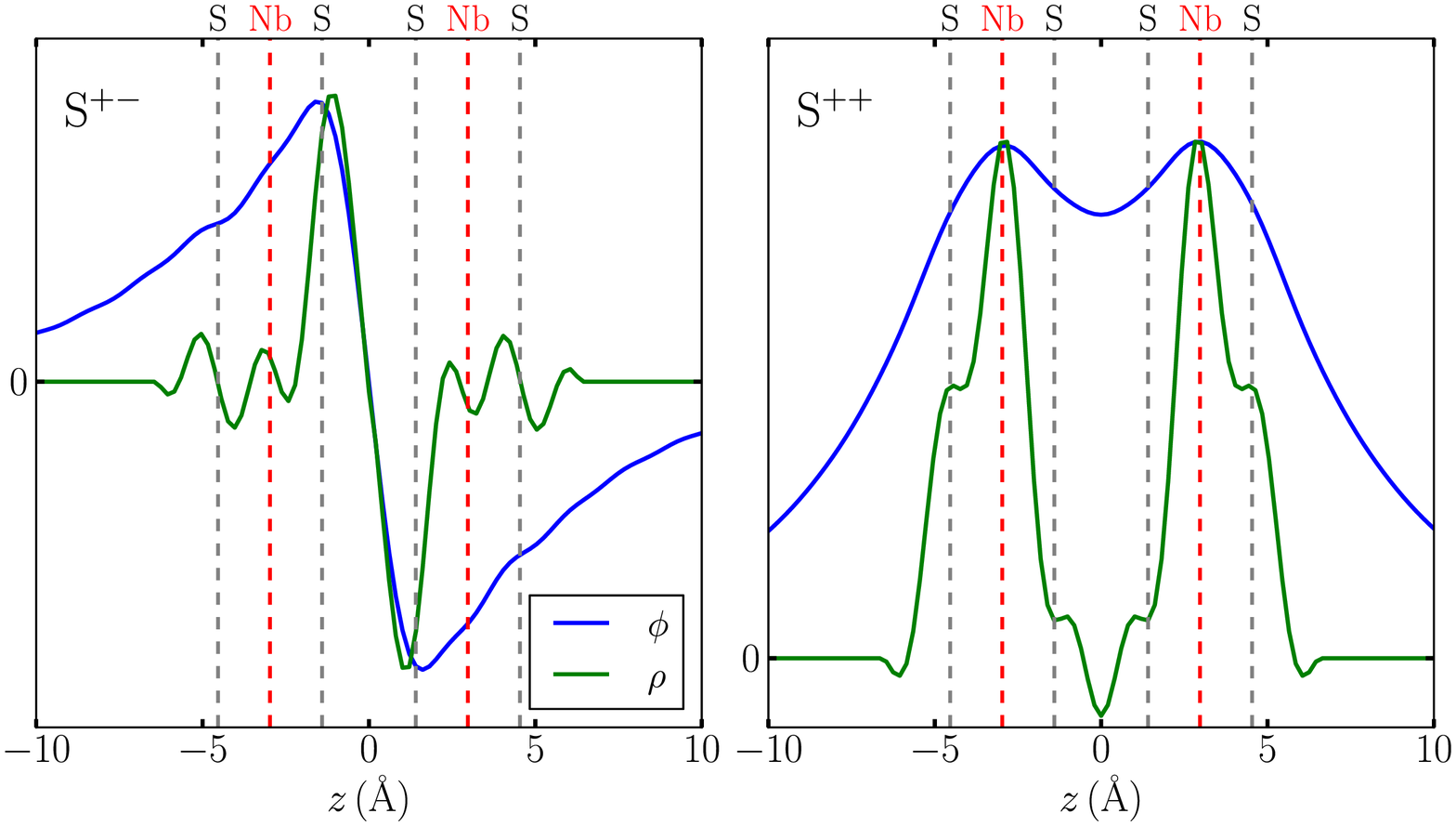}
   \includegraphics[height=0.28\linewidth]{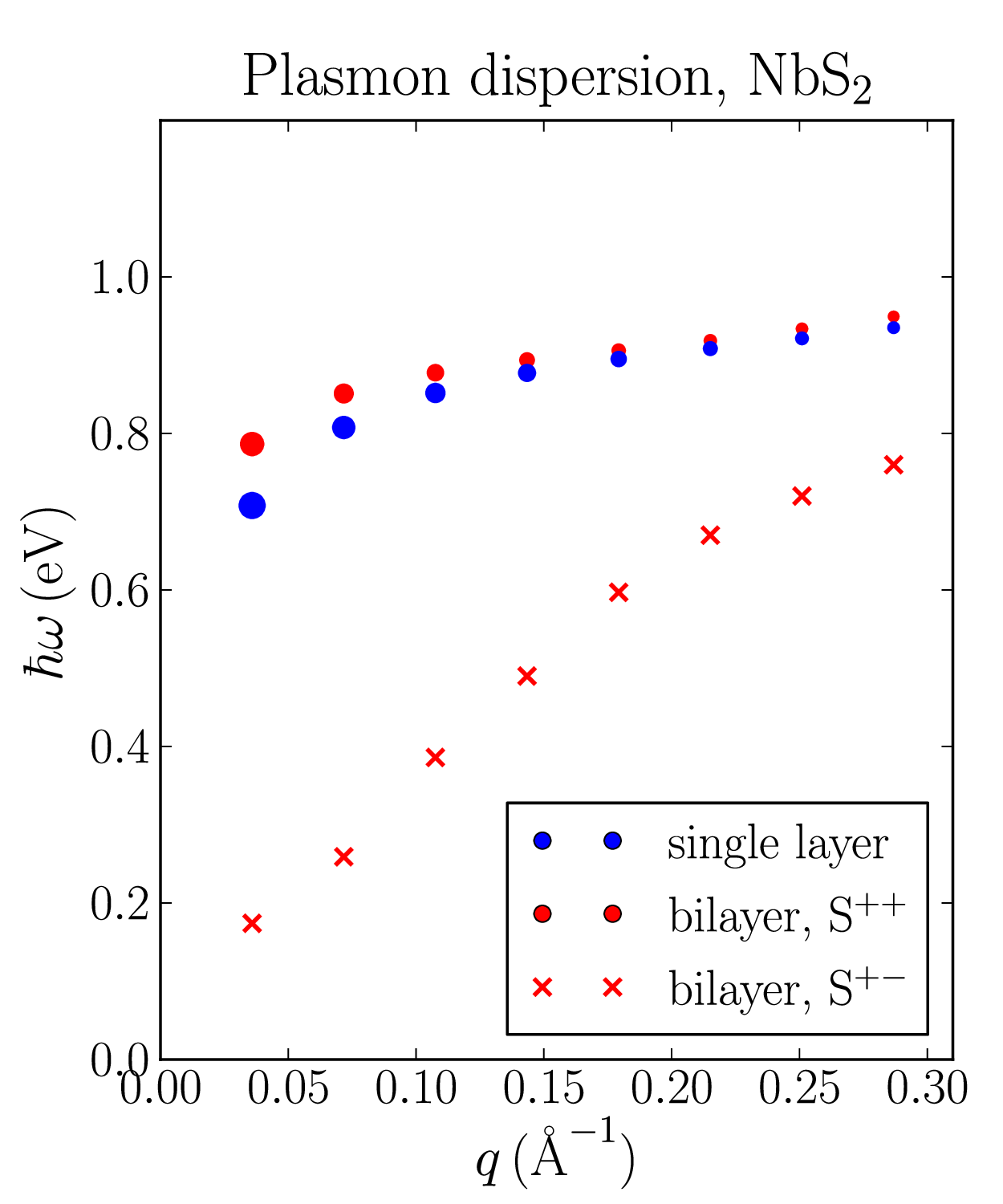}
\caption{Left panel: Eigenvalue loss spectrum for a bilayer NbS$_2$ structure. The imaginary part of the inverse eigenvalues of the dielectric matrix, $\epsilon_n(\omega)$, are plotted as function of frequency. The two dominating eigenvalue curves, highlighted by the blue and green curves, correspond to the symmetric and anti-symmetric plasmon modes, respectively. Middle panel: Spatial shape of the symmetric ($S^{++}$) and anti-symmetric ($S^{+-}$) plasmon modes plotted in the direction perpendicular to the layer. The charge density, $\rho$, and the corresponding potential, $\phi$, are shown in green and blue, respectively. Right panel: The energy dispersion of the two modes of the bilayer is shown together with the single-layer result for \NbS. The $S^{++}$ mode is slightly blue-shifted compared to the single layer result as expected, whereas the $S^{+-}$ mode is significantly redshifted, and approaches the $S^{++}$ mode with increasing $q$. 
}
\label{fig:bilayer}
\end{center}
\end{figure*} 

As seen in Fig.~\ref{fig:tase2} the inclusion of interband transitions
have a strong influence on the plasmons which are significantly
red-shifted and weakened. To define the borders of the intra- and
interband continua we have calculated the threshold energy where
$\mr{Im} \, \chi^{0, \mr{intra}}_{\mb{G},\mb{G}'}(\mb{q},\omega)$
drops to zero and $\mr{Im} \,\chi^{0,
  \mr{inter}}_{\mb{G},\mb{G}'}(\mb{q},\omega)$ starts to increase from
zero, respectively. The regions thus defined are indicated in the
figure. The upper edge of the intraband continuum is well separated
from the plasmon resonance, whereas the lower edge of the interband
continuum is at approximately 1 eV at low $q$, and descends to lower
energies as $q$ increases. At $q \approx 0.1$ \AA$^{-1}$ the plasmon
enters the dissipative interband region and thus acquires a finite
lifetime.  Returning to the band structures, the onset of the
interband continuum around 1 eV is seen to originate from transitions
between the occupied bands just below the Fermi level and the metallic
band. We note that transition from an undamped to a damped plasmon is
not very clear in Fig. \ref{fig:bands} where all the EELS peaks have a
finite width.  This is due to the numerical smearing which introduces
an artificial broadening of all spectral features. 
We mention that the effect of interband transitions on plasmons has been studied previously, e.g. for simple metals\cite{ferdi} and graphene\cite{hwang}. 

\subsection{Bilayer NbS$_2$}
Finally, we have studied a bilayer structure of NbS$_2$. The layers are hexagonally stacked and the separation
set equal to half the experimental lattice constant of the bulk
material ($c$ = 11.89 \AA).  For such a bilayer system, classical
plasmon hybridization theory predicts the existence of two plasmon
modes for the combined structure, corresponding to the symmetric and anti-symmetric, or bonding and anti-bonding, combinations of the uncoupled plasmon modes of the single layers\cite{Prodan_2003}. 

In order to investigate the spatial form of the plasmon modes of the
bilayer we follow a recently developed method.\cite{Andersen_2012}
As explained in Sec. \ref{sec.plasmon} the full dielectric matrix is diagonalized for each frequency
point to obtain its eigenvalues, $\epsilon_n(\mb{q},\omega)$ and left
and right eigenfunctions, $\phi_n(\mb{q},\omega)$ and
$\rho_n(\mb{q},\omega)$. The imaginary part of the inverse
eigenvalues, $-\mr{Im} \, \epsilon_n^{-1}(\mb{q},\omega)$, then provide a resolution the loss spectrum into independent
 dielectric eigenmodes.  Modes corresponding to distinct peaks in the eigenvalue
 loss spectrum can be identified as the plasmon modes of the system,
 and the corresponding left and right eigenfunctions represent the
 potential and charge density of the plasmon excitation, respectively.\cite{Andersen_2012}

 In the left panel of Fig.~\ref{fig:bilayer} the eigenvalue loss
 spectrum is shown for a momentum transfer of $q \approx 0.25$
 \AA$^{-1}$. The soectrum is dominated by two eigenvalue curves,
 indicated by green and blue solid lines. For comparison, the
 eigenvalue loss spectrum for the monolayer systems (not shown)
 reveals a single peak very similar to both the blue curve in Fig. 4
 (symmetric mode) and the EELS spectrum of the monolayer (shown in
 Fig. \ref{fig:bands} for TaSe$_2$). Returning to Fig.
 ~\ref{fig:bilayer}, the eigenvectors belonging to the two peaks
 reveal an anti-symmetric and symmetric mode, labelled $S^{+-}$ and
 $S^{++}$ respectively. The $S^{+-}$ mode corresponds to an out of
 phase charge oscillation located on the two $\mr{S}$ atoms in the
 interface.  The energy of this mode is significantly red shifted
 compared to the monolayer plasmon and shows a close to linear
 dispersion, see right panel of Fig.~\ref{fig:bilayer}. However, the
 strength of this mode is vanishing at low $q$, which is possibly
 related to the close proximity of the two layers. The $S^{++}$ mode
 can be characterized as an in-phase oscillation located on the
 transition metal atoms of the two layers. The induced density
 ($\rho$) of this mode within each layer is very similar to that found
 for the plasmon in monolayer NbSe$_2$ (not shown). The $S^{++}$ mode is
 blue shifted relative to the single-layer result, particularly in the
 small $q$ regime where the overlap between the potentials from the two
 layers is largest (recall the potential of a 2D charge density wave
 decays as $e^{-qz}$)\cite{ferrell}.

\section{Conclusion}
In conclusion, we have investigated the plasmonic properties of six
monolayers of metallic transition metal dichalchogenides using
first-principles calculations. Significant deviations from a 2D
free-electron model were found due to coupling to interband
transitions which redshift the plasmon energy by up to 1 eV. For
momentum transfers of $q\approx 0.1$ \AA$^{-1}$ the plasmon enters the
dissipative electron-hole continuum. For a bilayer NbSe$_2$ structure
we found two plasmon modes which can be regarded as symmetric and
anti-symmetric combinations of two unperturbed monolayer plasmons. The
anti-symmetric mode shows close to linear dispersion, but has very
small strength, while the dispersion of the symmetric mode follows the
monolayer plasmon with a slight blue shift.

\section{Acknowledgements}
KST acknowledges support from the Danish Council for Independent Research's Sapere Aude Program through grant no. 11-1051390. The Center for Nanostructured
Graphene (CNG) is sponsored by the Danish National Research Foundation, Project DNRF58.


\bibliographystyle{unsrt}

\end{document}